\documentclass[aps,prl,onecolumn,amsmath,amssymb,superscriptaddress]{revtex4}
\usepackage{graphicx}
\usepackage{amssymb}
\usepackage{natbib}
\usepackage{color}

\newcommand{\beq}{\begin{eqnarray}}
\newcommand{\eeq}{\end{eqnarray}}

\begin{document}

\title{Dynamic spin-lattice coupling and nematic fluctuations in NaFeAs}
\author{Yu Li}
\affiliation{Department of Physics and Astronomy,
Rice University, Houston, Texas 77005, USA}

\author{Zahra Yamani}
\affiliation{Canadian Nuclear Laboratories,
Chalk River Laboratories,Chalk River, Ontario KOJ 1JO, Canada}

\author{Yu Song}
\affiliation{Department of Physics and Astronomy,
Rice University, Houston, Texas 77005, USA}

\author{Weiyi Wang}
\affiliation{Department of Physics and Astronomy,
Rice University, Houston, Texas 77005, USA}

\author{Chenglin Zhang}
\affiliation{Department of Physics and Astronomy,
Rice University, Houston, Texas 77005, USA}

\author{David W. Tam}
\affiliation{Department of Physics and Astronomy,
Rice University, Houston, Texas 77005, USA}

\author{Tong Chen}
\affiliation{Department of Physics and Astronomy,
Rice University, Houston, Texas 77005, USA}

\author{Ding Hu}
\affiliation{Department of Physics and Astronomy,
Rice University, Houston, Texas 77005, USA}
\affiliation{Center for Advanced Quantum Studies and Department of Physics, Beijing Normal University, Beijing 100875, China}

\author{Zhuang Xu}
\affiliation{Center for Advanced Quantum Studies and Department of Physics, Beijing Normal University, Beijing 100875, China}

\author{Songxue Chi}
\affiliation{Quantum Condensed Matter Division, Oak Ridge National Laboratory, Oak
Ridge, Tennessee 37831, USA}

\author{Ke Xia}
\affiliation{Center for Advanced Quantum Studies and Department of Physics, Beijing Normal University, Beijing 100875, China}
\affiliation{Synergetic Innovation Center for Quantum Effects and Applications (SICQEA), Hunan Normal University, Changsha 410081, China}
\author{Li Zhang}
\affiliation{Center for Advanced Quantum Studies and Department of Physics, Beijing Normal University, Beijing 100875, China}

\author{Shifeng Cui}
\affiliation{Center for Advanced Quantum Studies and Department of Physics, Beijing Normal University, Beijing 100875, China}

\author{Wenan Guo}
\affiliation{Center for Advanced Quantum Studies and Department of Physics, Beijing Normal University, Beijing 100875, China}

\author{Ziming Fang}
\affiliation{Center for Advanced Quantum Studies and Department of Physics, Beijing Normal University, Beijing 100875, China}

\author{Yi Liu}
\affiliation{Center for Advanced Quantum Studies and Department of Physics, Beijing Normal University, Beijing 100875, China}

\author{Pengcheng Dai}
\email{pdai@rice.edu}
\affiliation{Department of Physics and Astronomy,
Rice University, Houston, Texas 77005, USA}
\affiliation{Center for Advanced Quantum Studies and Department of Physics, Beijing Normal University, Beijing 100875, China}

\date{\today}
\pacs{74.70.Xa, 75.30.Gw, 78.70.Nx}

\begin{abstract}
We use inelastic neutron scattering to study acoustic phonons and spin excitations in single crystals of NaFeAs, a parent compound of iron pnictide superconductors.
NaFeAs exhibits a tetragonal-to-orthorhombic structural transition at $T_s\approx 58$ K
and a collinear antiferromagnetic (AF) order at $T_N\approx 45$ K.  While longitudinal and out-of-plane transverse acoustic phonons behave as expected, the in-plane transverse acoustic phonons reveal considerable softening on
cooling to $T_s$, and then harden on approaching $T_N$ before saturating below $T_N$. In addition, we find that spin-spin correlation lengths of low-energy magnetic excitations within the FeAs layer and along the $c$-axis increase dramatically below $T_s$, and
show weak anomaly across $T_N$. These results suggest that the electronic nematic phase present in the paramagnetic tetragonal phase is closely associated with dynamic spin-lattice coupling, possibly arising from the one-phonon-two-magnon mechanism.
\end{abstract}

\maketitle

\section{Introduction}
Spin waves (excitations) and phonons are two fundamental quasiparticles in a solid describing propagating disturbance of the ordered magnetic moment and lattice vibrations, respectively \cite{Lovesey}. They are the byproducts of linearized theories that ignore all the higher-order terms than quadratic term and neglect quasiparticle-quasiparticle interactions \cite{landau} unlikely interacting with each other \cite{Zhitomirsky}. However, in systems with multiple coupled degrees of freedom, the interaction affects the self energy of the quasiparticles and changes their intrinsic dispersion relationship. Therefore, discovering and understanding how the interaction and coupling changes the spectra of spin excitations and phonons and further influences the electronic properties of solids are one of the most important themes in modern condensed matter physics.

In general, spin-lattice (magnon-phonon) coupling can modify the spin excitations in two different ways. First, the
static lattice distortion coupled with magnetic order, as seen in some parent compounds of iron-based superconductors \cite{PD_RMP,Inosov}, can dramatically change the effective magnetic exchange anisotropy in a near
square lattice of the antiferromagnetic (AF) ordered phase \cite{zhao09}. Second, the dynamic lattice vibrations (phonons) interacting with time-dependent spin excitations may create energy gaps in the magnon dispersion at the nominal intersections of the magnon
and phonon modes \cite{Lovesey}, as seen in insulating non-collinear antiferromagnet
(Y,Lu)MnO$_3$ \cite{YMnO3}. Similarly, spin waves can hybridize with optical phonon modes near
the zone boundary and form broadened magnon-phonon excitations in metallic ferromagnetic manganites \cite{dai00}.
In both cases, magnon-phonon interactions only form hybridized excitations at the magnon phonon intersection points
due to the repulsive magnon-phonon dispersion curves.

In this paper, we report a different form of
magnon-phonon interaction in NaFeAs,
which is
a parent compound of iron pnictide superconductor and exhibits a tetragonal-to-orthorhombic structural transition at $T_s\approx 58$ K followed by a collinear AF order at $T_N\approx 45$ K [Fig. 1(a)] \cite{Parker,slli09,park2012,Wright2012}.
For temperatures between $T_N$ and $T_s$, NaFeAs has an electronic nematic phase,
a form of electronic order which breaks the rotational symmetries
without changing the translational symmetry of the underlying lattice
\cite{fradkin,CFang,CXu,Fernandes2011,SLiang13,fisher,Kuo2016,MYi,YZhang2012,Allan2013,Rosenthal,Xingye,QZhang2015,WLZhang2016}, which
may play an important role in high temperature superconductivity \cite{Metlitski,Lederer}. Instead of the usual magnon-phonon interaction where the magnon dispersion would intersect with the phonon dispersion in reciprocal space, we find evidence for magnon-phonon interaction in NaFeAs, where the low-energy
acoustic phonons near the nuclear zone center $\Gamma$ point in NaFeAs can interact with the low-energy spin excitations near the AF zone center $M$ point at ${\bf Q_{AF}}$ in reciprocal space without direct phonon-spin excitation crossing [Fig. 1(b) and 1(c), Figs. 2-5]. Since both the in-plane transverse acoustic (IPTA) phonons [Fig. 1(b)] and spin excitations show dramatic
anomaly across $T_s$, our results indicate that
the magnon-phonon interaction may be responsible for the low-temperature electronic nematic phase in NaFeAs \cite{fradkin,CFang,CXu,Fernandes2011,SLiang13,fisher,Kuo2016,MYi,YZhang2012,Allan2013,Rosenthal,Xingye,QZhang2015,WLZhang2016}. Although not conclusively established, one possible
microscopic interpretation of these results is the one-phonon-two-magnon scattering mechanism
\cite{theory1,theory2}.
By considering dynamical spin exchange coupling tuned by local lattice vibration, we calculate
the effect of magnon-phonon coupling in one-dimensional (1D) spin chain.
Further phonon dispersion calculations using density function theory (DFT) on NaFeAs demonstrates that softening of the IPTA mode occurs at $q\rightarrow 0$ in the presence of stripe type AF spin-spin correlations, consistent with our observation.
Therefore, instead of a pure spin \cite{CFang,CXu,Fernandes2011,SLiang13} or lattice (orbital)
\cite{Kruger09,WLv09,CCLee,Kontani,Baek15,bohmer15} degree of freedom, dynamic spin-lattice coupling is
important for the
electronic nematic phase and anisotropic electronic properties in the paramagnetic state of NaFeAs.

\section{Results}
\subsection{Neutron Scattering}
We chose to study NaFeAs
because transport \cite{fisher,Kuo2016}, angle resolved photoemission spectroscopy (ARPES) \cite{MYi,YZhang2012},
scanning tunneling microscopy \cite{Allan2013,Rosenthal}, and neutron scattering experiments \cite{Xingye,QZhang2015,WLZhang2016} provided ample evidence for the presence of an electronic nematic phase in $T_N\leq T\leq T_s$.  By assuming weak spin-orbit and spin-lattice coupling, most theoretical understandings of the nematic phase involve pure ``spin'' or ``orbital'' degrees of freedom.
In the first case, the electronic nematic phase originates from spin fluctuations which couple quadratically to the nematicity but have no direct coupling to lattice \cite{CFang,CXu,Fernandes2011}. Here, key predictions of the theory, such as the enhanced spin excitation intensity and correlation length at the AF ordering wave vector ${\bf Q_{AF}}$
in the paramagnetic phase of iron pnictides \cite{Fernandes,Fernandes12}, have been confirmed by inelastic neutron scattering experiments \cite{Xingye,QZhang2015,WLZhang2016}.
Alternatively, ferro-orbital order or orbital fluctuations \cite{Kruger09,WLv09,CCLee,Kontani,Baek15,bohmer15}, seen in ARPES measurements as a pronounced
energy splitting of bands with $d_{xz}$ and $d_{yz}$ orbital characters above $T_N$ \cite{MYi,YZhang2012}, may
induce the electronic nematicity through the direct
ferro-orbital order (fluctuations) and orthorhombic lattice distortion (or vibration)
coupling \cite{bohmer16,XYLu2016}.

The difficulty in sorting out the microscopic origin of the nematic phase in iron pnictides lies in the fact that the relevant degrees of freedom, such as spin, lattice, and orbital are closely entangled \cite{bohmer16,XYLu2016}.
  However, DFT calculations of iron pnictides \cite{Yin1,Mazin1,Yildirim} shows
a remarkable sensitivity of superconductivity and magnetism to the lattice, particularly the iron pnictogen height,
thus suggesting a strong dynamic spin-lattice coupling in this system.
Measurements and calculations on the phonon dispersions in CaFe$_2$As$_2$ and BaFe$_2$As$_2$ suggest that magnetism must be taken into account when calculating the phonon frequencies and phonons are strongly coupled not to the static AF order, but to spin fluctuations \cite{Zbiri,CaFe2As2_phonon,Reznik1}.
In addition, inelastic X-ray and neutron scattering measurements on AF ordered BaFe$_2$As$_2$ and SrFe$_2$As$_2$ where $T_N\approx T_s$
show a  softening of the IPTA phonon on cooling to $T_s$ [Fig. 1(a) and 1(b)], followed by a dramatic hardening
below $T_N/T_s$ \cite{Jennifer,Parshall}. For electron-doped superconducting Ba(Fe$_{1-x}$Co$_x$)$_2$As$_2$ with $T_c<T_N<T_s$, the softening of the IPTA phonon stops at $T_s$ on cooling, and the hardening of the mode passes through $T_N$ unabated before softening again below $T_c$ \cite{weber}. On the other hand, a sharp enhancement in the in-plane
spin-spin correlation length was found below $T_s$ at ${\bf Q_{AF}}$ in LaFeAsO and Ba(Fe$_{1-x}$Co$_x$)$_2$As$_2$ \cite{QZhang2015}, suggesting
a strong feedback effect of the structural transition and nematic order on the low-energy spin fluctuations. Although these results suggest a correlation between the evolution of the transverse acoustic phonon and low-energy spin fluctuations, the measurements were carried out on different samples by different groups, thus making it difficult to establish a direct connection between lattice vibrations and spin fluctuations.

\begin{figure}[t]
\includegraphics[scale=.25]{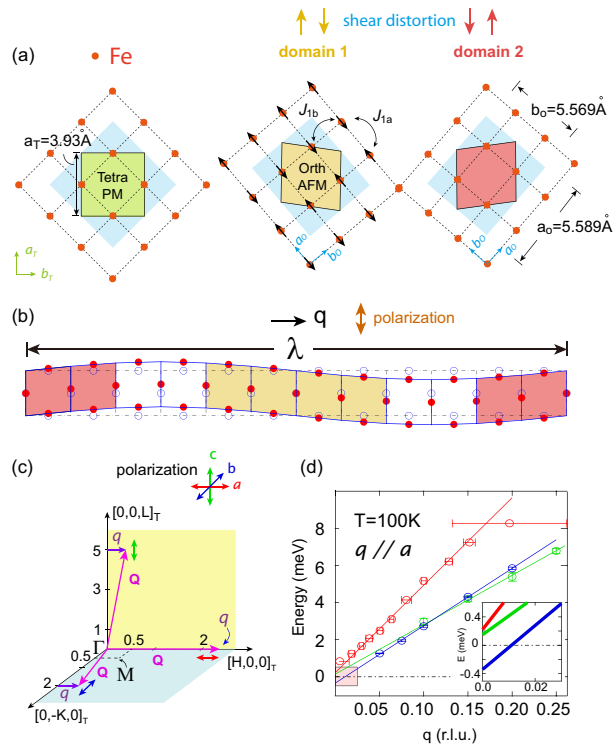}
\caption{(Color online) (a) Left: schematics of the Fe layers in NaFeAs. Red spots represent Fe atoms and the tetragonal (orthorhombic) unit cell is shown in green (blue) shaded square. Middle and right panels are the two orthorhombic domains. The yellow and red area are the tetragonal unit cells with opposite
shear distortions. The arrows in the middle panel indicates spin direction in the AF ordered phase. The orthorhombic lattice parameters are $a_o=5.589$ \AA\ and
$b_o=5.569$ \AA. (b) A particular IPTA phonon mode with a momentum propagating horizontally to the right but with Fe atoms oscillating vertically.
$\lambda$, the size of the probing domain, is inversely related to $q$. (c) The three different equivalent positions in reciprocal space used in our neutron scattering measurements in order to determine
all three acoustic phonon branches. The purple arrows with $q$ are the momentum of the measured phonons along the $[H,0,0]$ direction and the double-headed arrows
represent the phonon polarizations along the $a$ (red, LA), $b$ (blue, IPTA) and $c$ (green, OPTA) axis.
(d) The corresponding dispersions of the three phonon modes at 100 K estimated from data in \cite{SI}. The inset shows expanded
view of the low-energy part of the dispersion.}
\end{figure}

To resolve this problem, we use inelastic neutron scattering to study phonons and spin excitations in single crystals of NaFeAs, which is the parent compound of
NaFe$_{1-x}$Co$_x$As family of iron-based superconductors \cite{GTTan}.
Our measurements were carried out on the C5 triple-axis spectrometer at the Canadian Neutron Beam center, Chalk River Laboratories,
 with a fixed final neutron energies of
$E_f$=8.04 meV and 14.56 meV for phonon measurements and $E_f$=13.7 meV for spin fluctuation measurements. Some
scans of the IPTA phonon were carried on the HB-3 triple-axis spectrometer at the High Flux Isotope Reactor (HFIR), Oak Ridge National Laboratory,
with $E_f$=14.7 meV.
The wave vector transfer ${\bf Q}$ in three-dimensional (3D) reciprocal space in {\AA}$^{-1}$ is defined as ${\bf Q}=H{\bf a^{*}}+K{\bf b^{*}}+L{\bf c^{*}}$,
with ${\bf a^{*}}=\frac{2\pi}{a}{\bf \hat{a}}$, ${\bf b^{*}}=\frac{2\pi}{b}{\bf \hat{b}}$
and ${\bf c^{*}}=\frac{2\pi}{c}{\bf \hat{c}}$, where $H$, $K$ and
$L$ are Miller indices. In the paramagnetic tetragonal state, we have $a=b=a_T=b_T=3.93$ \AA\ and $c=6.98$ \AA\ [Fig. 1(a)].
In the AF ordered phase, $a_o\approx b_o \approx \sqrt{2}a$ \AA\ [Figs. 1(b) and 1(c)] and AF spin fluctuations occur at ${\bf Q_{AF}}\approx (0.5,0.5)$ \cite{PD_RMP}.
The phonon measurements were made on a single piece of NaFeAs single crystal weighing $\sim$2.5 g with mosaic $<1^\circ$ aligned in
the $[H,K,0]$ and $[H,0,L]$ scattering planes [Fig. 1(c)]. The spin fluctuation measurements were carried out using
a well aligned pack of single crystals ($\sim$10 g) with mosaic $< 3^{\circ}$ aligned in the $[H,H,L]$ scattering plane [Fig. 3(a)].

\begin{figure}[t]
\includegraphics[scale=.40]{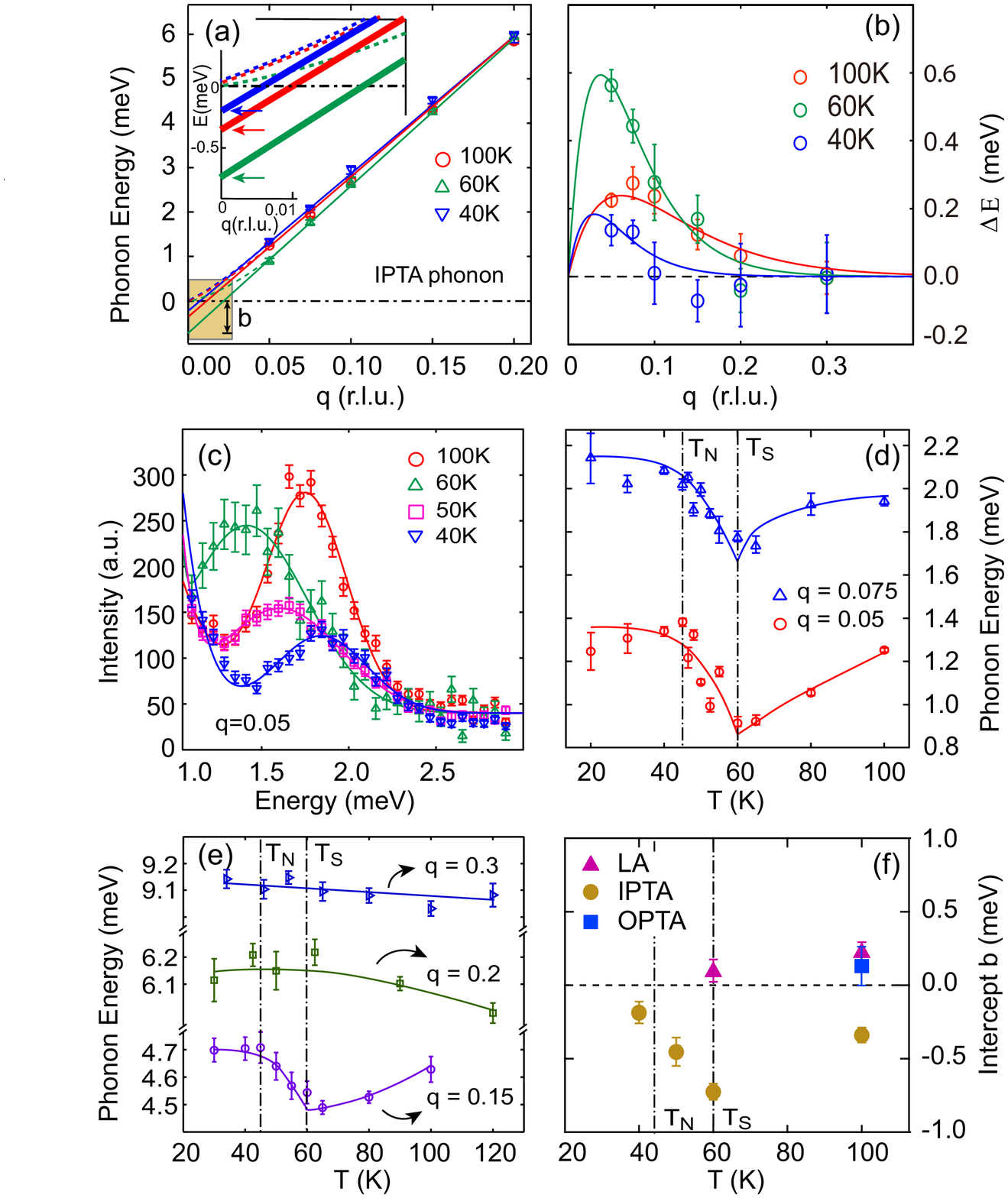}
\caption{(Color online) (a) Temperature dependence of the phonon dispersions of the IPTA mode
near the zone center at 100 K, 60 K, and 40 K. The inset shows the expanded view of the low-energy part of the dispersions.
(b) The estimated amplitude of the softening of the IPTA phonons at difference temperatures obtained by subtracting an assumed linear dispersion $E=vq$. The solid lines are guides to the eye.
(c) Temperature dependence of constant-$q$ scans at $q=0.05$ for the IPTA mode. (d) and (e) Temperature dependence of the phonon energy measured at $q= 0.05,0.075$ and $0.1, 0.15$,respectively. The phonon softening on cooling to $T_s$ is observed at $q=0.05$, 0.075, and 0.15 rlu
but absent at $q=0.20$ and 0.3 rlu.
The phonon hardening below $T_s$ are seen for wave vectors below $q=0.2$ r.l.u.
(f) Temperature dependence of the intercept $b$ estimated for all
three acoustic branches shown in Fig. 1(d). }
\end{figure}

To measure IPTA, out-of-plane transverse acoustic (OPTA) and longitudinal acoustic (LA)
phonon modes in NaFeAs, we probed wave vectors ${\bf Q}=(q,2,0)$,
$(q,0,5)$, and $(2+q,0,0)$, respectively, where ${\bf q}$ is the reduced momentum transfer away from the $\Gamma$ point
in reciprocal space [Fig. 1(c)] \cite{Jennifer,Parshall}.
In the long wave length (small $|{\bf q}|$) limit, one would expect linear dispersions for acoustic phonons $E=vq$, where $E$ is the
energy of the mode and $v$ is the sound velocity. Figure 1(d) shows our measured dispersions of LA, OPTA, and IPTA acoustic phonon modes at 100 K \cite{SI}.
While LA and OPTA phonons behave as expected and passed through the zero energy at $q=0$, the IPTA mode,
 which corresponds to the $q=0$ shear modulus mode
$C_{66}$ seen in the ultra sound spectroscopy measurements \cite{Fernandes10,c66_japan,bohmer}, has a lower sound velocity as $q\rightarrow 0$
[ dashed lines in Fig.2(a)]. The solid lines are linear fits to the large wave vector data.

Figure 2(a) shows dispersions of the IPTA phonons determined from a series of constant-${\bf Q}$ scans at different temperatures \cite{SI}. While the high-$q$ ($> 0.2$ r.l.u.) part of the dispersion hardens slightly with decreasing temperature
due to stiffer lattice at low temperature (but showed no anomaly across $T_s$ and $T_N$),
the low-$q$ ($<0.2$ r.l.u.) part of the dispersion is strongly temperature dependent [Fig. 2(b)]  \cite{SI}.
This is seen in the temperature dependence of the constant-${\bf Q}$ scans at $q=0.05$ r.l.u. [Fig. 2(c)], where
the peak center shifts from 1.25 meV at 100 K to 0.9 meV at 60 K near $T_s$. On further cooling to below $T_s$, the phonon hardens with decreasing temperature. The temperature dependence of the phonon energy at different $q$ is summarized
Figs. 2(d) and 2(e).

In general, acoustic phonons with different polarizations should exhibit similar linear dispersion dependence at similar temperatures.
  When phonons couple with electrons or other elementary excitations, their peak position, width, and intensity might change \cite{CDW}.
While LA and OPTA phonon dispersions in Fig. 1(d) follow $E=vq$ below $q \leq 0.2$ r.l.u., IPTA phonon dispersion is $E=vq+b$ with $b<0$ [Fig. 1(d)].
Since IPTA phonon at $q=0.2$ r.l.u. is weakly temperature dependent [Fig. 2(a) and 2(b)], its temperature dependent effect
for $q<0.2$ r.l.u. will be absorbed into the parameter $b$ [Fig. 2(a)] \cite{SI}. Figure 2(f) shows temperature dependence
of $b$ for the LA, OPTA, and IPTA phonons obtained by fitting with $E=vq+b$.  While the LA and OPTA phonons behave as expected with $b=0$,
 the values of $b$ for the IPTA mode are negative in both the AF ordered and paramagnetic state, thus suggesting
the presence of the IPTA phonon-elementary excitation (either spin or orbital) coupling that is independent of the magnetic ordering. Since such coupling is absent
for the LA and OPTA phonon modes, we conclude that the nonlinear dispersion curves seen in the IPTA phonon mode at small $q$
and their strong temperature dependence around $T_s$ are due to the IPTA phonon and elementary (spin or orbital) excitation coupling
with the absolute value $b$ reflecting the coupling strength.

\begin{figure}[t]
\includegraphics[scale=.60]{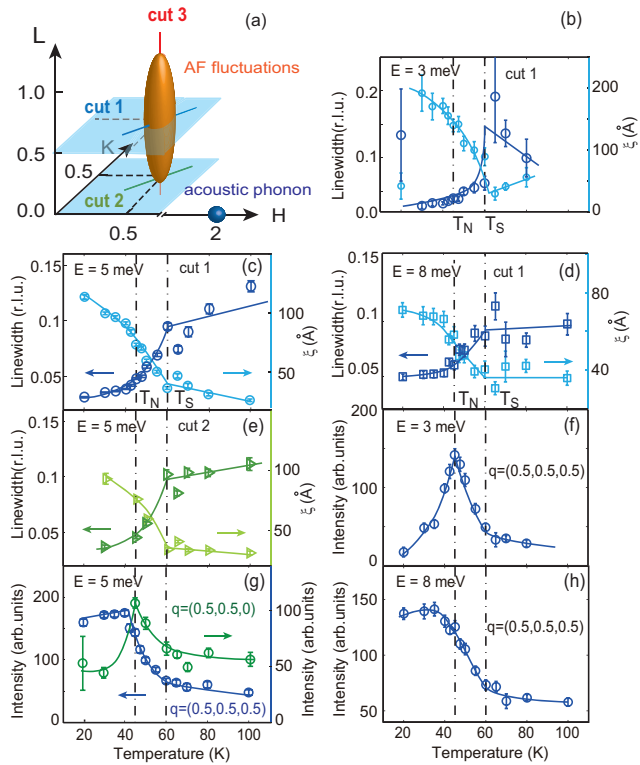}
\caption{(Color online) (a) Sketch of the reciprocal space in which AF spin fluctuations and acoustic phonons are marked as yellow ellipsoid and blue sphere. The directions along which we measured in the experiment are shown as blue, green and red lines. (b-e)Temperature dependent line widths and correlation lengths of the low energy spin fluctuations. They are measured along the cut 1 direction at (b) 3 meV, (c) 5 meV, (d) and 8 meV, (e) along the cut 2 direction at 5 meV. (f-h) Temperature dependence of the peak intensity in each scan. It is clear that there are negative correlations between the peak intensity and the peak width
before the system enters into the AF ordered state, representing a redistribution of spectral
weight in the nematic phase below $T_s$.
The vertical dashed lines mark the magnetic and structural
transition temperatures.
}
\end{figure}

\begin{figure}[t]
\includegraphics[scale=.60]{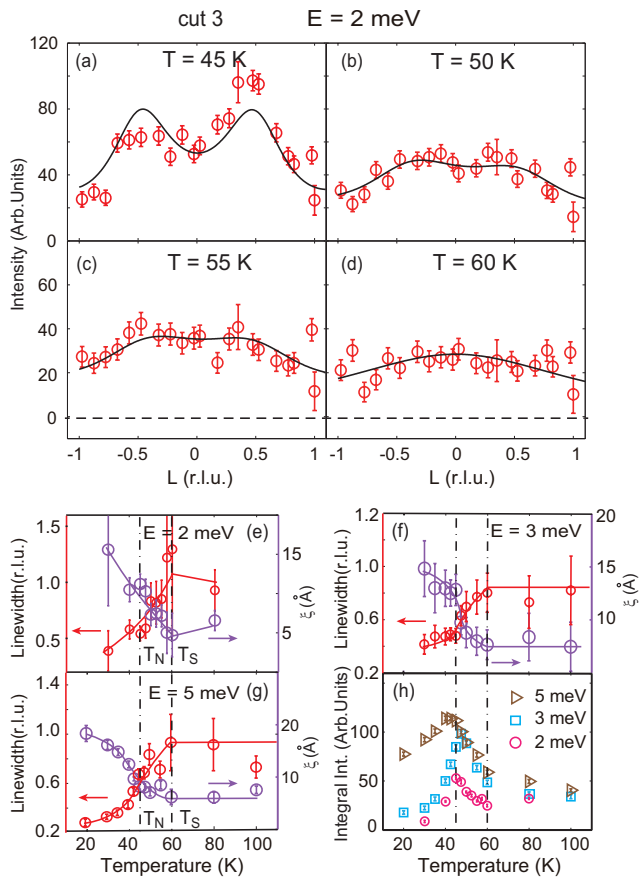}
\caption{(Color online) (a-d) Constant-$E$ ($E=2$ meV) scans along the cut 3 direction as shown in Fig. 3(a) at 45 K, 50 K, 55 K and 60 K.
Background scans are obtained by measuring at $q=(0.4,0.4,L)$ and $(0.6,0.6,L)$ and subtracted from the raw data.
The scan at 60 K is featureless. On cooling to 55 K, two weak peaks show up around $L = \pm 0.5$ rlu.
Upon further cooling, these peaks become prominent at $T=45$ K. They disappear below $T_N$ due to the opening of a spin gap.
The total magnetic scattering intensity increases on cooling. The solid lines are fits with periodic
Lorentz function \cite{SI}.
 (e-g)Temperature dependence of the line widths and correlation lengths of the low energy spin fluctuations
along the $c$-axis direction [cut 3 in Fig. 3(a)] at 2 meV, 3 meV, and 5 meV.
(h) Temperature dependence of the integrated intensity along the cut 3 direction at different energies \cite{SI}.
}
\end{figure}

\begin{figure}[t]
\includegraphics[scale=.18]{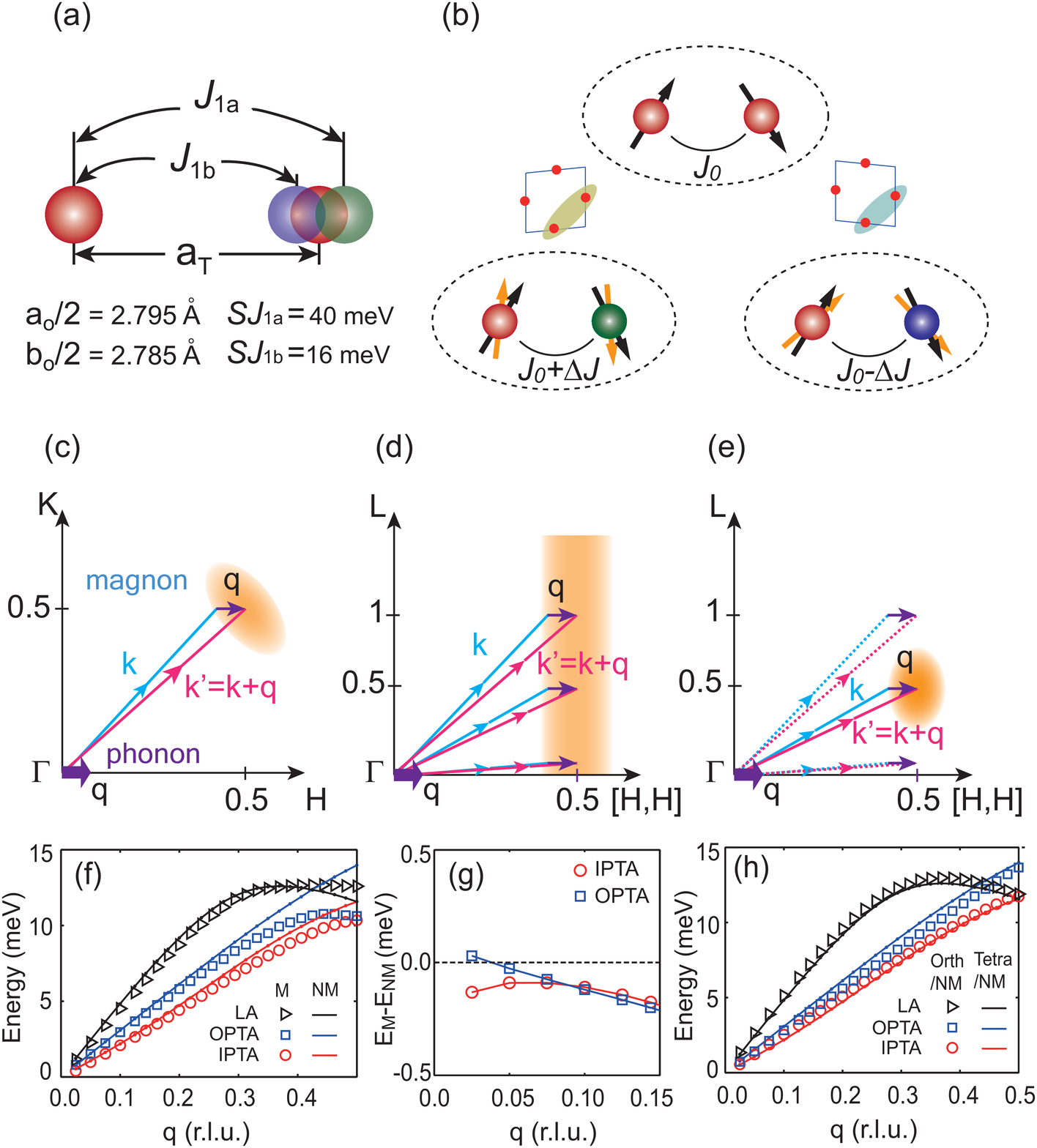}
\caption{(Color online) (a) Schematics of the nearest neighbor Fe-Fe bond. The small difference between $a_o/2$ and $b_o/2$ and
the large difference between $J_{1a}$ and $J_{1b} $ renders a nonnegligible ${\Delta J}/{\Delta r}$, which entangles the lattice and magnetism.
(b) One pair of magnetic Fe ions in the paramagnetic tetragonal phase with the magnetic binding energy of $E_b=J_0 S_1S_2\cos\theta$. As the Fe-Fe bond stretches, the spin exchange coupling increases and thus there is an energy transfer from the lattice to the spin system. (c) The scattering triangle required by the momentum conservation in the dynamic spin-lattice coupling process within the $[H,K]$ plane. (d) and (e) The momentum triangles in the $[H,H,L]$ plane above and below $T_s$. As spin fluctuations undergoes a 2D-to-3D transition, a large part of
the spin-lattice coupling scattering channel would be blocked.
(f) Comparison of the phonon dispersions obtained from DFT calculations with (markers, $M$) and without (solid lines, $NM$)
magnetism assuming a tetragonal lattice. We note that nonmagnetic tetragonal phase
does not really exist in NaFeAs since spin correlations are strong at temperatures
far above $T_N$ and $T_s$. (g) Wave vector dependence of $E_M-E_{NM}$ for IPTA and OPTA phonon modes.
(h) Comparison of the phonon dispersions obtained from DFT calculations with nonmagnetic ($NM$) Fe site assuming
 orthorhombic (markers) and tetragonal (solid lines) lattice. The black, blue, and red color represents
LA, OPTA, and IPTA phonon modes.
}
\end{figure}

In addition to phonons,
spin fluctuations are another important elementary excitations in iron pnictides \cite{fradkin,PD_RMP}. Distinct from
acoustic phonons which are observed near nuclear Bragg peak positions such as $(2,0,0)$, low-energy spin fluctuations in NaFeAs
occur near the ${\bf Q_{AF}}=(0.5,0.5,0.5)$ AF zone center in reciprocal
space [Fig. 3(a)] \cite{PD_RMP}. Figure 3 summarizes temperature dependence of the in-plane spin-spin correlations for
spin fluctuations with energies of $E=3$ meV, 5 meV, and 8 meV along the cut 1 and cut 2 directions as shown in Fig. 3(a).  Figure 3(b-e) shows temperature dependence of the line width
and the corresponding spin correlation length $\xi$, respectively.  A sharp reduction in the peak linewidth and a dramatic increase in the spin-spin correlation lengths are seen below $T_s$ \cite{SI}. They are accompanied by the increase of the intensities at the peak centers, reflecting a redistribution of the magnetic spectral weight in the nematic phase.
While these results are somewhat different  from a neutron scattering work on NaFeAs \cite{JJLiu}, which finds no dramatic anomaly across $T_s$, they are consistent with previous work on LaFeAsO and Ba(Fe$_{0.953}$Co$_{0.047}$)$_2$As$_2$,
which was interpreted as resulting from the long-range nematic order below $T_s$ \cite{QZhang2015}.
We note that a phonon at $q=0.05$ probes coherent lattice vibrations of $\sim$20 unit cells,
giving a dynamic length scale of $\sim$40-80 \AA . These values are similar to the observed spin-spin correlation lengths above $T_s$,
suggesting that correlated spin domains might be limited by sizes of the dynamic structural domains.

To further understand spin excitations across $T_s$, we carefully studied temperature dependence of spin-spin correlations along the $c$-axis.
Figures 4(a)-(d) are the background subtracted wave vector scans along the $[0.5,0.5,L]$ direction at spin excitation energy of $E=2$ meV and different temperatures. At $T=60$ K ($> T_s$), the scattering is essentially featureless and decreases with increasing $L$ due to the iron magnetic
form factor, suggesting that magnetic excitations are two-dimensional (2D) rods along the $c$-axis in reciprocal space [Fig. 4(d)].
On cooling to $T=55$ K ($<T_s$), Two broad peaks emerge around $L=\pm 0.5$ rlu indicating increased $c$-axis spin-spin correlation [Fig. 4(c)].
Upon further cooling to
$T=50$ K and 45 K ($>T_N$), these peaks become progressively better established [Fig. 4(b) and 4(a)].
The solid lines of Fig. 4(a)-(d) are fits to these peaks with a periodic Lorentzian function \cite{SI}.
Figures 4(e)-(g) plot temperature dependence of the fitted width and corresponding
spin correlation length at $E= 2$ meV, 3 meV, and 5 meV. Inspection of the Figures reveals a clear change of
the $c$-axis spin correlation length below $T_s$, suggesting that a spin fluctuation
crossover from 2D to 3D occurs below $T_s$ and above $T_N$.  Figure 4(h) shows temperature dependence
 of the integrated intensities over the range from $L=-1$ to $L=1$, which again reveal an clear increase in magnetic scattering below $T_s$.
From results described in Figs. 3 and 4, we conclude that the low energy spin fluctuations in NaFeAs respond dramatically
to the structural transition and change its spectral distribution below $T_s$.

\subsection{Theoretical considerations and density functional theory calculation}
	
Given the simultaneous occurrence of the dramatic increasing spin correlation length at ${\bf Q_{AF}}$  and a hardening of the IPTA phonon at $\Gamma$ point below $T_s$ (Figs. 2-4), it is interesting to ask if the low-energy spin fluctuations are coupled to the IPTA phonons.
To address this question, we consider the differences between paramagnetic tetragonal
and orthorhombic phases. In the paramagnetic tetragonal phase, iron spins are completely random, spending equal amount
of time along all directions. On cooling to paramagnetic orthorhombic phase below $T_s$, the direction of the spin alignment
gets approximately fixed to the low-temperature AF ordered collinear phase due to dynamic
spin-lattice coupling,
so that within a structural domain iron spins in the nearest
neighbors are predominantly anti-parallel along the orthorhombic $a_o$ axis and parallel along the $b_o$ axis. As a consequence, the magnetic
correlation length changes abruptly across $T_s$ but does not become infinite due to the lack of long-range magnetic order.
Regardless of a detailed microscopic spin-lattice coupling mechanism, this picture is valid and suggests a magnetic origin of the nematic phase \cite{Fernandes2011}.

To provide a possible microscopic description of the spin-lattice coupling, we consider the magnitude of the effective magnetic exchange couplings in NaFeAs.  By fitting the spin wave spectra throughout the Brillouin zone
at 5 K using a Heisenberg Hamiltonian, it was found that the effective magnetic exchange couplings along the orthorhombic $a_o$ and $b_o$ axes are
$SJ_{1a}=40\pm 0.8$ meV and $SJ_{1b}=16\pm 0.6$ meV, respectively, where $S\approx 1$ is the spin of the system \cite{CLZhang2015}.
Since the distances between the nearest Fe-Fe neighbors in the AF orthorhombic phase are $a_o/2=2.795$ \AA\ and $b_o/2=2.785$ \AA\ \cite{slli09},
the large changes in the magnetic exchange in the AF ordered phase may arise from tiny differences in the in-plane Fe-Fe distances
within the Heisenberg Hamiltonian [Fig. 5(a)] \cite{Bi}.
In the paramagnetic tetragonal state, the shear modulations of the IPTA phonon mode [Figs. 1(a) and 1(b)]
may induce large changes in magnetic exchange through dynamic spin-lattice coupling [Fig. 5(b)]. In previous ultrafast optical spectroscopy measurements on Ba(Fe$_{1-x}$Co$_x$)$_2$As$_2$, it was found
that coherent optical phonon motions of As atoms on ultrafast timescales induced by a femtosecond optical pulse can produce transient dynamic AF and/or nematic
order at temperatures above $T_s$ through possible modifications of the Fermi surfaces and dynamic spin-lattice coupling \cite{Ultrafast1,Ultrafast2}.
For conventional AF insulator such as (Y,Lu)MnO$_3$, the effect of dynamic spin-lattice coupling is to
create energy gaps in the magnon dispersion at the nominal intersections of the magnon and phonon modes \cite{YMnO3}.
However, this cannot explain our observation in NaFeAs, since the low-energy acoustic phonons and spin fluctuations
are located at different wave vectors in reciprocal space [Fig. 3(a)].

To solve this problem, we consider the one-phonon-two-magnon scattering mechanism \cite{theory1,theory2}.
Although such mechanism is no different from the usual magnetoelastic coupling or magnon-phonon coupling, it provides a simple pictorial way to visualize such coupling.
Theoretically, it has been found that
spin-lattice coupling due to the one-phonon-two-magnon scattering process in an antiferromagnet
can affect the spin wave broadening/damping, dispersion, and
intensity  \cite{ZJXiong}.
Since spin fluctuations in iron pnictides
exhibit spin-wave-like features above $T_N$ up to very high temperatures \cite{Leland}, such magnon-phonon coupling must also exist in the paramagnetic tetragonal phase. Figure 5(b) schematically
illustrates how such a dynamic spin-lattice scattering mechanism might work.
In the paramagnetic tetragonal phase with strong AF spin fluctuations, the neighboring spins are aligned
anti-parallel on a time scale determined by the energy of
the fluctuations and the average moment of the system at elastic position ($E=0$ or infinite time) is still zero.
When an IPTA phonon with momentum ${\bf q}$ and energy $E_0$ is emitted near $\Gamma$ point [Fig. 1(b)], the Fe-Fe distance stretches and thus the exchange coupling increases
from $J_0$ to $J_0 + \Delta J$ [Fig. 5(b)] \cite{SI}. This represents an energy transfer from the lattice to the spin system, annihilating one phonon
and producing spin fluctuations, and vice versa.

To test the possible spin-lattice coupling, we built a toy model of 1D spin chain in which the spin exchange coupling is tuned by the local lattice vibration induced by a passing phonon (see Ref. \cite{SI}, section \uppercase\expandafter{\romannumeral4}).
Similar to the recent calculations on the effect of one-phonon-two-magnon to the indirect $K$-edge resonant inelastic x-ray scattering spectrum of
a 2D Heisenberg antiferromagnet \cite{ZJXiong}, we find that spin waves in 1D chain are also affected by the lattice distortion.  In particular, the periodic lattice distortion affect spin waves, which in turn
leads to the phonon softening at small $q$ similar to our observation in NaFeAs (Fig. S21). Since analytical analysis in the 2D case is difficult, we
preformed DFT calculations to compare the phonon dispersions of NaFeAs with and without magnetic moment on the Fe site.
The open triangles, squares, and circles in Figure 5(f) are DFT calculated LA, OPTA, and IPTA phonon dispersions
with Fe magnetic moment, respectively.  The solid lines in the same figure show identical calculations without magnetism.
While the main effect of magnetism is to soften the OPTA and IPTA phonons near the zone boundary ($q=0.5$) consistent
 with earlier work \cite{Zbiri,CaFe2As2_phonon,Reznik1},
its effect on the low-energy OPTA and IPTA phonons can be seen by comparing phonon energies with magnetism ($E_M$) and without magnetism ($E_{NM}$).
Figure 5(g) shows wave vector dependence of $E_M-E_{NM}$, revealing a clear difference between OPTA and IPTA phonons below
$q=0.075$.  These calculations suggest that the presence of magnetic moment on Fe has much larger effect on the low energy
IPTA phonons, consistent with our observation of the phonon softening determined by the intercept $b$ [Fig. 2(b)].
To test if the small orthorhombic lattice distortion in the nematic phase of NaFaAs can affect acoustic phonons without magnetism, we show
in Fig. 5(h) the calculated phonon dispersions in orthorhombic and tetragonal states in the nonmagnetic
state. Within the errors of our calculation, we find no evidence of phonon softening
due to lattice orthorhombicity. From these results, we conclude that
the presence of magnetism and spin-spin correlations selectively influence
the dispersion of the IPTA phonon at q$\rightarrow 0$.

\section{Discussion and conclusion}

Assuming that the spin-lattice coupling in NaFeAs is due to
 the one-phonon-two-magnon mechanism,
the scattering process must satisfy the energy and momentum conservation \cite{theory2,Dixon}.
Two magnons with momentum ${\bf k}$ and ${\bf k'}$ and energy $E$ and $E'$ with a relationship ${\bf q} = {\bf k} + {\bf k'}$ and $E_0 = E + E'$ (Positive $E$ means generating and negative $E$ means annihilating)
must be introduced as shown in Fig. 5(c). The scattering amplitude is constrained by the momentum triangle and depends on the detail of the spin fluctuation dispersion \cite{theory2}. In the paramagnetic tetragonal state, the magnetic scattering is a rod centered around ${\bf Q_{AF}}=(0.5,0.5)$ in the FeAs plane [Figs. 3(a), 4(d), and 5(d)]. This implies a large phase space for the dynamic spin-lattice coupling to occur.
However, when a 2D-to-3D crossover in spin fluctuations
occurs below $T_s$, the spin-lattice coupling phase space is greatly reduced as shown in Fig. 5(e).  Since the IPTA phonon mode
dynamically changes the nearest neighbor Fe-Fe distance which
is directly coupled to the nearest neighbor magnetic exchange coupling, the reduced scattering phase space below $T_s$ explains the
recovery of the phonon softening (\emph{phonon hardening}) at low temperature in NaFeAs.
Therefore, the observed ITPA phonon hardening below $T_s$ is a sign of reduced phonon-magnon coupling induced by the 2D-to-3D spin fluctuation crossover. In this picture, the narrowing of the low-energy in-plane spin fluctuations below $T_s$ is associated with the orthorhombic lattice distortion and nematic order, which increases the effective exchange
coupling along $a$ and therefore the spin wave velocity.
Since finite magnetic exchange coupling along the $c$-axis is required to establish the eventual 3D AF long-range order, the presence of the quasi-2D spin fluctuations in the paramagnetic tetragonal phase and the resulting large dynamic spin-lattice coupling are
directly associated with the observed nematic fluctuations in a variety of experiments
\cite{Kuo2016,MYi,YZhang2012,Allan2013,Rosenthal,Xingye,QZhang2015,WLZhang2016}.  The energy scale of the phonon softening $\sim$1 meV \cite{note2} is consistent with the slow recovery of the nematic order on a timescale of several or decades of ps observed by time-resolved polarimetry \cite{Ultrafast2}, further confirming its close relationship with the nematic fluctuations.

Although the one-phonon-two-magnon mechanism described above may provide a pictorial way to explain the experimentally
observed phonon and spin excitation anomaly across $T_s$ in NaFeAs, we recognize that any effect of dynamic spin-lattice coupling can be described as such process, and we have not established a
quantitative perturbative theory based on such process.
Regardless the microscopic origin of the observed phonon and spin excitation anomaly across $T_s$, our experimental results
revealed a clear coupling between
low-energy acoustic phonon and spin excitations at $T_s$.
Therefore, instead of  the pure spin or lattice (with associated orbital ordering) as the driving force for
the electronic nematic phase in iron pnictides, a strong dynamic spin-lattice-orbital coupling in the paramagnetic tetragonal phase
 is responsible for the observed electronic nematic order and anisotropic behavior.

\section{Acknowledgments}

We thank Rong Yu and Qimiao Si for helpful discussions.
The single crystal growth and neutron scattering work at Rice is supported by the
U.S. DOE, BES under contract no. DE-SC0012311 (P.D.).
A part of the materials work at Rice is supported by the Robert A. Welch Foundation Grants No. C-1839 (P.D.). The DFT calculation was partly supported by the National Natural Science Foundation of China (Grants No. 61604013) and the Fundamental Research Funds for the Central Universities (Grants No. 2016NT10).

\end{document}